\documentclass{article}
\usepackage[english]{babel}
\usepackage[utf8]{inputenc}
\usepackage{johd}

\usepackage{amsmath}
\usepackage{amsfonts}
\usepackage{graphicx}
\usepackage{ifoddpage}

\usepackage[margin=1in]{geometry}
\usepackage{setspace}
\usepackage{xcolor}

\title{MUTUALLY UNBIASED BASES IN 3 and 4 DIMENSIONS SEMI-QUANTUM KEY DISTRIBUTION PROTOCOL}

\author{Hasnaa Hajji$^{a}$\footnote{Email: \url{hasnaa_hajji@um5.ac.ma}} ,Morad El Baz$^{a}$\footnote{Email: \url{morad.elbaz@um5.ac.ma}} \\
        \small $^{a}$ ESMaR, Faculty of Sciences, Mohammed V University Rabat, Morocco.
}
\date{} 
\begin{document}

\maketitle

\begin{abstract} 
\noindent Semi-quantum key distribution is traditionally based on two-level quantum systems. In this paper, an unconditional security of a semi quantum key distribution protocol based on higher-dimensional systems using various mutually unbiased bases is presented. We first consider the three dimensional case using three and four mutually unbiased bases and derive a lower bound for the key rate as a function of the quantum channel’s noise. We then generalize the result to a semi-quantum key distribution protocol that employs different number of mutually unbiased bases for four-dimensional states. It is found that basing the semi-quantum key distribution protocol on higher-dimensional mutually unbiased bases can increase the tolerable threshold of the noise and the maximum achievable value of the secret key rate.  \end{abstract}

\noindent\\

%\noindent\authorroles{For determining author roles, please use following taxonomy: \url{https://casrai.org/credit/}. Please list the roles for each author.} 

\section{Introduction}
Mutually unbiased bases, originally introduced by Schwinger in 1960 \cite{Schwinger570} as optimum incompatible measurement bases, attracted considerable attention as a crucial resource for quantum techniques. Two bases $\{| v_i\rangle \}$ and $\{| w_j\rangle\}$, of a \textit{d}-dimensional Hilbert space $\mathcal{H}=\mathbb{C}^d$, are said to be \textit{mutually unbiased} if all inner products across their elements have the same magnitude: $|\langle v_i|w_j \rangle|=1/\sqrt{d}$ for all $i,j=1, 2 \dots, d$. It is well known that the maximum number of \textit{mutually unbiased bases} can be at most $d+1$ in a prime-dimension system or prime-power-dimension system \cite{Ivonovic_1981,WOOTTERS1989363}.

Mutually unbiased bases, usually abbreviated as MUBs, constitute an important cornerstone in contemporary quantum information processing. They are being used in quantum state determination\cite{Ivonovic_1981}, quantum state tomography \cite{PhysRevA.83.052332} as well as quantum error correction codes \cite{PhysRevLett.78.405}. In addition, MUBs have been successfully applied in quantum key distribution protocols, due to the fact that measurements in one basis preclude knowledge of the state in any of the others \cite{Bra,PhysRevLett.85.3313,PhysRevA.59.4238}.

On the other hand, high-dimensional systems have attracted considerable attention with many theoretical proposals and experimental realizations \cite{peresbook}. They offer advantages ranging from withstanding high channel noise levels, to possible implementation in the fundamental tests of quantum mechanics. High dimensional quantum key distribution based on qudit encoding ($d$-dimensional quantum systems) is an efficient technique to enhance the security of a quantum key distribution system against eavesdropping. Indeed, such protocols were shown to increase the channel capacity \cite{highdimqkd}.
 
 Already, conventional quantum key distribution (QKD) protocols based on the MUBs exhibit higher error rate and higher generation rates of secure key rate. 

Recently, the novel concept of “semi-quantumness” has been of great interest in the field of quantum cryptography and generally allows quantum information processing tasks to be accomplished using as few as possible quantum resources. Along this direction, a new class of quantum key distribution protocols was introduced \cite{1} whereby one of the two communication participants (typically the receiver) is limited to performing certain "classical" or "semi-quantum" operations. Namely, the receiver is only allowed to work directly with the computational basis. Protocols operating along these guidlines are known as semiquantum key distribution (SQKD) protocols. After the BKM07 protocol, suggested originally in 2007 by M. Boyer, D. Kenigsberg, and T.Mor in \cite{1}, several semi-quantum key distribution protocols appeared in the literature \cite{PhysRevA.79.032341,PhysRevA.83.046301,krawec2020security,hajji2021qutrit,PhysRevA.79.052312,5}.

 Since the use of high-dimensional MUBs for fully QKD protocols provides several benefits especially in key
generation rates, the question arises whether SQKD protocols using higher-dimensional MUBs can also result in higher generation rates of secure key as well as the amount of noise. In this paper, we address this question by focusing on "semi-quantumness" protocol in which we use three- and four-dimensional systems and vary the number of mutually unbiased bases used. We analyze each protocol in terms of key rate, considering the collective eavesdropping attacks. This allows us to compare the robustness against the eavesdropping of the various protocols.

The paper is organized as follows:
In the next section, we present a security analysis of a SQKD protocol \cite{hajji2021qutrit}, against collective eavesdropping strategy for a set of three-dimensional states using different numbers of mutually unbiased bases (instead of the conventional studies using just two bases). In section 3 we generalize the analysis to the interesting situations that can occur depending on whether we consider a subset of two, three, or four of these unbiased bases or all five mutually unbiased bases with four-dimensional quantum states. Brief concluding remarks are given at the end of the paper.

\section{COLLECTIVE EAVESDROPPING WITH \\ THREE-DIMENSIONAL QUANTUM STATES}
In this section we prove the security against collective eavesdropping strategies in semiquantum cryptographic protocols based on the transmission of three-dimensional systems (qutrits) with three and four mutually unbiased bases. We consider the general scenario where an eavesdropper performs a global measurement over an ensemble of stored ancilla, once the classical post-processing has taken place. Typically, a given eavesdropping strategy in a two way communication scheme can be characterized by two unitary operations. $U_F$ attacking the quantum systems as they go from Alice to Bob\footnote{ We use the general paradigm whereby Alice is considered the sender and Bob the receiver} \textit{i.e.} the forward direction and $U_R$ as they go back from Bob to Alice \textit{i.e.} the reverse direction.

Under collective attacks, we may employ the Devetak-Winter \cite{8} key rate equation using reverse reconciliation: 
\begin{equation}
\label{keyrateinf}
    r=inf[S(B|E)-H(B|A)],
\end{equation}
which states that the key rate is the difference between Eve's uncertainty on Bob's raw key, $S(B|E)$ (which, for a successful protocol,  should be high) and Alice's uncertainty of Bob's key, $H(B|A)=H(B,A)-H(A)$ (which should be low). The infimum is taking over all collective attacks that induce the observed error rates. Our main focus in this work is to determine a lower bound on the key rate of the protocol introduced in \cite{1}.

 Conventionally, the first basis of the protocol corresponds to the computational basis;\textit{e.g.} in a three-dimensional Hilbert space we denote it by $\{ |0\rangle, |1\rangle, |2\rangle\}$. Then, the most general eavesdropping strategy for qutrits, in the forward channel, is of the form:
\begin{eqnarray}
\label{UF3}
|0\rangle\otimes |E\rangle &\xrightarrow{U_F}&|0\rangle |e_{00}\rangle +|1\rangle |e_{01}\rangle+|2\rangle |e_{02}\rangle,  \nonumber \\
|1\rangle\otimes |E\rangle &\xrightarrow{U_F}& |0\rangle |e_{10}\rangle+|1\rangle |e_{11}\rangle+|2\rangle |e_{12}\rangle , \label{uf1} \\
|2\rangle\otimes |E\rangle &\xrightarrow{U_F}& |0\rangle |e_{20}\rangle+|1\rangle |e_{21}\rangle+|2\rangle |e_{22}\rangle . \nonumber
\end{eqnarray}
Then, on the backward channel, it will yield the final global state: 
\begin{equation}
\label{ur3}
    |i, e_{ji}\rangle \xrightarrow{U_R} |0, e^0_{i,ji} \rangle + |1, e^1_{i,ji}\rangle+ |2, e^2_{i,ji}\rangle,
\end{equation}
where Eve's initial state is denoted $|E\rangle$ and the states $|e_{ji}\rangle$ and $ |e^k_{i,ji} \rangle $ that are not necessarily normalized nor orthogonal, are her states after interacting on the forward channel and reverse channel respectively, with $i,j=0, 1, 2$.

The unitarity of $U_F$ imposes, on the scalar products between Eve’s output states, relations of the form:

\begin{eqnarray}
\langle e_{00} |e_{10} \rangle +\langle e_{01} |e_{11} \rangle +\langle e_{02} |e_{12} \rangle &=0\nonumber \\
\langle e_{10} |e_{20} \rangle +\langle e_{11} |e_{21} \rangle +\langle e_{12} |e_{22} \rangle &=0 \\
 \langle e_{00} |e_{20} \rangle +\langle e_{01} |e_{21} \rangle +\langle e_{02} |e_{22} \rangle &=0. \nonumber
\end{eqnarray}

Similar relations are imposed on the vectors $ |e^k_{i,ji} \rangle $ due to the unitarity of $U_R$.

Following Eve's attack, and conditioning on an iteration being used to contribute towards the raw key (\textit{i.e.} assuming that Alice sends a computational basis state, Bob  measures and re-sends then Alice measures in the computational basis), Alice and Bob may estimate the  probability  $p_{i,j,k}$, that Alice sends a state $|i \rangle$, Bob measures it as $|j \rangle$, then Alice finds it to be in the state $|k \rangle$ after receiving it back from Bob. These probabilities can be used to estimate the values $\langle e^a_{b,c}| e^a_{b,c} \rangle$. The inner products can be computed by observing the computational basis noise in the channel modeled as a ternary channel with parameter $Q$ \cite{hajji2021qutrit}.

Following equation (\ref{keyrateinf}), one can get a lower bound on the key rate by bounding the von Neumann Entropy $S(B|E)$. Owing to its strong sub-additivity, the von Neumann entropy for any tripartite system obeys
 \begin{equation}
 \label{strongsubadditivity}
 S(B|E)\geq S(B|EC) 
\end{equation}
where  $C$  is an additional random variable  introduced to form a four-party composite system $ABEC$. Inequality (\ref{strongsubadditivity}), thus supplies us with a lower bound on the key rate: 

 \begin{eqnarray}
 \label{strongsubadditivitykey rate}
r \geq inf [S(B|E)-H(B|A)]\geq inf [S(B|EC)-H(B|A)],
\end{eqnarray}
The system, $C$, we append is in a four-dimensional space, spanned by $\{|c,0\rangle, |c,1\rangle, |w,1\rangle, |w,2\rangle \}$ where $|c,i \rangle \langle c, i|$ is the event that Alice and Bob's raw key bits match (\textit{i.e.} are correct), and that the qutrit sent from Alice was flipped $i$ times, while $|w,i \rangle \langle w, i|$ denotes the event where their raw key bits don't match (\textit{i.e.} are wrong) while being flipped $i$ times also.

Incorporating this system, we can write the global state system $\rho_{BEC}$ as  a diagonal matrix, where the diagonal entries are elements of the form $\frac{1}{3} \langle e^k_{i,ji}|e^k_{i,ji}\rangle $ for all $|e^k_{i,ji}\rangle$, we readily compute the von Neumann entropy$S(B|EC)= S(BEC)-S(EC)$ to bound the final key rate $r$ .

Firstly, we can figure out that 
\begin{eqnarray}
\label{SBEC}
    S(BEC)=S(\rho_{BEC})=H(\frac{1}{3} p_{0,0,0},\frac{1}{3} p_{0,0,1},..., \frac{1}{3} p_{2,2,2}),
\end{eqnarray}a quantity that Alice and Bob may compute after parameter estimation. From this, computing  $S(EC)$  is trivial, by tracing out Bob's state we can get the state $ \rho_{EC}$ as

\begin{equation}
     \rho_{EC}  =(\frac{1}{3} t_1 \Tilde{\sigma_1}) \otimes |c,0 \rangle \langle c, 0| + (\frac{1}{3} t_2 \Tilde{\sigma_2}) \otimes |c,1 \rangle \langle c, 1| + (\frac{1}{3} t_3 \Tilde{\sigma_3}) \otimes |w,1 \rangle \langle w, 1| + (\frac{1}{3} t_4 \Tilde{\sigma_4}) \otimes |w,2 \rangle \langle w, 2|,
\end{equation}
where $\Tilde{\sigma_j}=\frac{\sigma_j}{t_j}$ is the normalized form of the positive semi-definite operators $\sigma_j$:
\begin{eqnarray}
\sigma_1 &=&|e^0_{0,0}\rangle \langle e^0_{0,0}|+| e^1_{1,4} \rangle \langle  e^1_{1,4}|+ |e^2_{2,8} \rangle \langle  e^2_{2,8}|\nonumber \\
 \sigma_2&=&|e^0_{0,3}\rangle \langle e^0_{0,3}|+|e^0_{0,6}\rangle \langle e^0_{0,6}|+|e^1_{1,1} \rangle \langle e^1_{1,1}|+| e^1_{1,7} \rangle \langle  e^1_{1,7}|+| e^2_{2,2} \rangle \langle  e^2_{2,2}|+| e^2_{2,5} \rangle \langle  e^2_{2,5}|\nonumber \\
 \sigma_3&=&| e^1_{0,0} \rangle \langle  e^1_{0,0}|+|e^2_{0,0} \rangle \langle  e^2_{0,0}|+|e^0_{1,4}\rangle \langle e^0_{1,4}|+| e^2_{1,4} \rangle \langle  e^2_{1,4}|+|e^0_{2,8}\rangle \langle e^0_{2,8}|+| e^1_{2,8} \rangle \langle  e^1_{2,8}|\nonumber \\
 \sigma_4&=&|e^1_{0,3} \rangle\langle e^1_{0,3}|+| e^2_{0,3} \rangle \langle e^2_{0,3}|+| e^1_{0,6} \rangle \langle  e^1_{0,6}|+| e^2_{0,6} \rangle \langle  e^2_{0,6}|+|e^0_{1,1}\rangle \langle e^0_{1,1}|+| e^2_{1,1} \rangle \langle e^2_{1,1}| \nonumber \\
 &&+|e^0_{1,7}\rangle \langle e^0_{1,7}|+| e^2_{1,7} \rangle \langle  e^2_{1,7}|+|e^0_{2,2}\rangle \langle e^0_{2,2}|+| e^1_{2,2} \rangle \langle  e^1_{2,2}|+|e^0_{2,5}\rangle \langle e^0_{2,5}|+| e^1_{2,5} \rangle \langle  e^1_{2,5}|.  \nonumber
\end{eqnarray}
We  have defined $t_j = Tr{(\sigma_j)} > 0$, for $j = 1, 2, 3, 4$ with $t_1$ representing the total probability that there is no error between Alice and Bob in both channels (forward and backward), $t_2$ the total probability that there is an error in the forward channel, $t_3$ the total probability that there is an error in the backward channel and $t_4$ is the total probability that there is an error in both channels (forward and backward).

Therefore, we can bound  $S(EC)$ as 
\begin{equation}
\label{sec}
    S(EC)=S(\rho_{EC})= H(\frac{1}{3} t_1,...,\frac{1}{3}t_4)+ \frac{1}{3}\sum^4_{j=1} t_j S(\Tilde{\sigma_j}).
\end{equation} 

A lower bound on the key rate equation (\ref{keyrateinf}), requires an upper bound on $S(EC)$, which we can find easily using (\ref{sec}):
\begin{equation}
\begin{split}
    S(EC)\leq& H\left(\frac{1}{3}t_1,\frac{1}{3}t_2,\frac{1}{3}t_3, \frac{1}{3}t_4\right) \\
    &+\frac{1}{3}(t_2+t_3+t_4)+\frac{1}{3}t_1 S(\Tilde{\sigma_1}).
\end{split}
\end{equation}
Evidently if the noise of the quantum channel is low, then $p_{i,j,k}$ should be low except for $p_{0,0,0}$, $p_{1,1,1}$ and $p_{2,2,2}$ that should be high. Thus, we can get the lower bound on the key rate $r$ by finding an upper bound on $ S(\Tilde{\sigma_1})$.\\
After some algebraic manipulation, the eigenvalues of $\Tilde{\sigma_1}$, denoted $\Tilde{\lambda_0}$, $\Tilde{\lambda_1}$ and $\Tilde{\lambda_2}$ can be found:

\setlength{\arraycolsep}{0.1em}
\begin{eqnarray}
\Tilde{\lambda_0}&=& 0, \\
\Tilde{\lambda_1}&=&\frac{1}{2}+\frac{\sqrt{4p+p_{0,0,0}^2-2 p_{0,0,0}p_{1,1,1}+p_{1,1,1}^2-2 p_{0,0,0}p_{2,2,2}-2 p_{1,1,1}p_{2,2,2}+p_{2,2,2}^2}}{2(p_{0,0,0}+p_{1,1,1}+p_{2,2,2})}, \label{lambda1tilde} \\
\Tilde{\lambda_2}&=&\frac{1}{2}-\frac{\sqrt{4p+p_{0,0,0}^2-2 p_{0,0,0}p_{1,1,1}+p_{1,1,1}^2-2 p_{0,0,0}p_{2,2,2}-2 p_{1,1,1}p_{2,2,2}+p_{2,2,2}^2}}{2(p_{0,0,0}+p_{1,1,1}+p_{2,2,2})},\label{lambda2tilde}\\
\nonumber
\end{eqnarray}
where $p=\left(|\langle e^0_{0,00}|e^1_{1,11}\rangle|^2+|\langle e^0_{0,00}|e^2_{2,22}\rangle|^2+|\langle e^1_{1,11}|e^2_{2,22}\rangle|^2\right)$.

Incorporating everything together yields the following upper bound on $S(EC)$
\begin{equation}
\begin{split}
\label{SEC}
    S(EC)\leq& H\left(\frac{1}{3}t_1,\frac{1}{3}t_2,\frac{1}{3}t_3, \frac{1}{3}t_4\right)\\  +&\frac{1}{3}(t_2+t_3+t_4)+\frac{1}{3}t_1\left(H(\Tilde{\lambda_1})+H(\Tilde{\lambda_2})\right).
\end{split}
\end{equation}
Applying equations (\ref{strongsubadditivitykey rate}),(\ref{SBEC}),(\ref{SEC}), the key rate bound is found to be
\begin{align}
\label{keyratebound3}
    r &\geq H(\frac{1}{3} p_{0,0,0},\frac{1}{3} p_{0,0,1},..., \frac{1}{3} p_{2,2,2})- H\left(\frac{1}{3}t_1,\frac{1}{3}t_2,\frac{1}{3}t_3, \frac{1}{3}t_4\right) \nonumber \\
    &-\frac{1}{3}(t_2+t_3+t_4)-\frac{1}{3}t_1 H(\Tilde{\lambda_1},\Tilde{\lambda_2} )\\
    &+H(p_A(0),p_A(1),p_A(2)) - H(\left\{ p(i,j)\right\}_{i,j=0,1,2}  ), \nonumber 
\end{align}
with $p_A(a)$ being the probability that A's raw key bit is $a$, $p(i,j)$ for $\left\{i,j \right\}= \left\{0, 1, 2, 3 \right\}$ the probability that Bob's raw key is $i$ while Alice's is $j$ and we have used the notation
\begin{align}
    H(\left\{ p(i,j)\right\}_{i,j=0,1,2})=&
   H(p(0,0),p(0,1),p(0,2), \nonumber \\
   &p(1,0), p(1,1),p(1,2),  \\
   &p(2,0),p(2,1),p(2,2)) \nonumber.
\end{align}

From the above inequality,  the  eigenvalues (\ref{lambda1tilde}) and (\ref{lambda2tilde}) depend on the values $p_{i,j,k}$ but also on the quantity $p$ which cannot be directly observed. However, by using the error rate in the different numbers of mutually unbiased bases,  Alice and Bob may determine bounds on these quantities. We carry this task in the following subsections in the case of a protocol using, respectively, three and four MUBs.

\subsection{Three mutually unbiased bases}
The quantity  $ p=(|\langle e^0_{0,00}|e^1_{1,11}\rangle|^2+|\langle e^0_{0,00}|e^2_{2,22}\rangle|^2+|\langle e^1_{1,11}|e^2_{2,22}\rangle|^2)$ can be bound by considering the noise in two mutually unbiased bases, given by 
 \begin{align}
\label{Xbasis}
    |X \rangle_0 &=\frac{1}{\sqrt{3}}(|0\rangle+ |1\rangle+ |2\rangle),\nonumber \\ 
    |X\rangle_1 &=\frac{1}{\sqrt{3}}(|0\rangle+\eta |1\rangle+\eta^{*}|2\rangle),\\
    |X\rangle_2 &=\frac{1}{\sqrt{3}}(|0\rangle+\eta^{*}|1\rangle+\eta|2\rangle).\nonumber
\end{align}
\begin{align}
    \label{Ybasis}
    |Y \rangle_0 &=\frac{1}{\sqrt{3}}(\eta |0\rangle+ |1\rangle+ |2\rangle),\nonumber \\ 
    |Y \rangle_1 &=\frac{1}{\sqrt{3}}(|0\rangle+\eta |1\rangle+|2\rangle),\\
    |Y \rangle_2 &=\frac{1}{\sqrt{3}}(|0\rangle+|1\rangle+\eta|2\rangle).\nonumber
\end{align}
with $\eta=e^{\frac{2i\pi}{3}}$. It is easy to check that the scalar product between any basis states belonging to different bases is $\frac{1}{\sqrt{3}}$.
%Any basis vectors These two bases are mutually unbiased since $\langle$ =1/	3 with i , j=0,1,2.

We focus on the process that Alice choose to encode her qutrit in a state from the 6 states defined above while Bob chooses to reflect, and Alice measures in the same basis she originally used to prepare it. Since Bob chooses to reflect, the two way quantum channel becomes, essentially, a one way channel with Eve attacking through the unitary operator $V=U_RU_F$. Assuming, without loss of generality, that Eve's ancilla is cleared to the zero state $|0\rangle_E$, her action on the basis states can be described as follows
\begin{equation}
    |i\rangle\otimes |0\rangle_{E} \xrightarrow{V} |i\rangle | f_{ii} \rangle + |i+1\rangle | f_{ii+1}\rangle+ |i+2\rangle | f_{ii+2}\rangle.
\end{equation}
The unitarity of V leads to the constraints
\begin{equation}
\label{UniV3}
    \langle f_{ii} |f_{ji}\rangle+ \langle f_{ij} |f_{jj}\rangle+ \langle f_{ik} |f_{jk}\rangle=0.
\end{equation}
where $i=0, j=1, k=2$ and cyclic permutations of these values. Moreover, it has been shown \cite{Cirac} that the symmetry reduces considerably the complexity of the analysis. The symmetry condition is defined by imposing some restrictions on the scalar products which characterize the unitary operation $V$ of Eve’s eavesdropping strategy. Mainly, that the scalar products of Eve’s eavesdropping strategy should be invariant under the exchange of the indices (0, 1, and 2). Therefore it is possible to divide the scalar products into six different groups:  
\begin{align}
\label{sym3d}
a= &\langle f_{ii}|f_{ij}\rangle, &&\text{for $i\neq j$} ,\nonumber  \\
b= &\langle f_{ii}|f_{jk}\rangle, &&\text{where}\;  i,j, \text{and}\; k \; \text{are all different,}\nonumber \\
c= &\langle f_{ij}|f_{ik}\rangle,  &&\text{where}\;  i,j, \text{and}\; k \; \text{are all different,}\nonumber \\
z= &\langle f_{ij}|f_{ji}\rangle,  &&\text{for $i\neq j$} ,\\
m= &\langle f_{ij}|f_{ki}\rangle,  &&\text{where}\;  i,j, \text{and}\; k \; \text{are all different,}\nonumber \\
t= &\langle f_{ii}|f_{jj}\rangle,  &&\text{for $i\neq j$; $t$ is real.} \nonumber 
\end{align}
 Taking into account the unitarity (\ref{UniV3}) and symmetry conditions (\ref{sym3d}), we derive the following expression
\begin{align}
t&=1-\frac{1}{4}(P_{X_0X_1}+P_{X_0X_2}+P_{X_1X_0}+P_{X_1X_2} \nonumber \\
& \quad+P_{X_2X_1}+P_{X_2X_0}+P_{Y_0Y_1}+P_{Y_0Y_2}+P_{Y_1Y_0}  \\
& \quad +P_{Y_1Y_2}+P_{Y_2Y_1}+P_{Y_2Y_0})-\frac{1}{2} Re(m). \nonumber
\end{align}
where a probability $P_{ij}$ for $\left\{i,j \right\}= \left\{X_0, X_1, X_2\right\}$ or $\left\{i,j \right\}=\left\{Y_0, Y_1, Y_2\right\}$, is that of Alice measuring the returned qutrit in the state $|j\rangle$ when she originally prepared it in the state $|i\rangle$.

From the Cauchy-Schwarz's inequality, it follows that the quantity $ X_{3MUBs}=Re(p)$ can bounded as follows 

\begin{equation}
\label{X3MUBS}
\begin{split}
    X_{3MUBs}  &\geq 3-\frac{3}{4}(P_{X_0X_1}+P_{X_0X_2}+P_{X_1X_0}+P_{X_1X_2}\\
   & \quad +P_{X_2X_1}+P_{X_2X_0} +P_{Y_0Y_1}+P_{Y_0Y_2}+P_{Y_1Y_0}\\
   & \quad +P_{Y_1Y_2}+P_{Y_2Y_1}+P_{Y_2Y_0})\\
 & \quad -\frac{3}{2}(\sqrt{p_{001}p_{102}} + \sqrt{p_{011}p_{102}} + \sqrt{p_{021}p_{102}} \\
 & \quad + \sqrt{p_{001}p_{112}}+ \sqrt{p_{011}p_{112}} + \sqrt{p_{021}p_{112}} \\
& \quad +  \sqrt{p_{001}p_{122}} + \sqrt{p_{011}p_{122}} + \sqrt{p_{021}p_{122}})\\
& \quad -3(\sqrt {p_{000} p_{101}}+\sqrt {p_{010} p_{101}}+\sqrt {p_{020} p_{101}} \\
& \quad +\sqrt {p_{010} p_{111}}+\sqrt {p_{020} p_{111}}+\sqrt {p_{000} p_{121}}\\
& \quad +\sqrt {p_{010} p_{121}}+\sqrt {p_{020} p_{121}}).
\end{split}
\end{equation}

This result is plotted in Figure \ref{fig:3D} together with the other cases of two MUBs \cite{hajji2021qutrit} and four MUBs to be derived in the next subsection. We considered two scenarios. Namely, the dependent channel, where the noise on both the forward and reverse channel are depending on each other, and the independent channel, where these noises are independent of each other. For the later, the $\left\{ |k\rangle \right\}_{i}$ for ${k=X,Y,Z},{i=0,1,2} $ basis noise accumulated by the qutrit as it travel through both channels when Bob chooses to reflect is $Q_{ind}=2Q(2-3Q)$. In the dependent case the corresponding noise is simply $Q_{dep}=Q$. For three MUBs, the key rate is positive as long as $Q_{\textit{indep}} \leq 3.95\%$ for the independent channel case and $Q_{\textit{dep}} \leq 6.89\%$ for the dependent channel case. For any value of the noise smaller than this threshold value, Alice and Bob may distill a secure secret key.

\subsection{Four mutually unbiased bases}
In this part, we express a lower bound of the quantity $p=\left(|\langle e^0_{0,00}|e^1_{1,11}\rangle|^2+|\langle e^0_{0,00}|e^2_{2,22}\rangle|^2+|\langle e^1_{1,11}|e^2_{2,22}\rangle|^2\right)$ by using three mutually unbiased bases rather than two mutually unbiased bases.  

In addition to the two bases defined before (\ref{Xbasis}, \ref{Ybasis}), we define the third basis by substituting in the later $\eta$ with $\eta^*$ :
\begin{align}
    |Z \rangle_0 &=\frac{1}{\sqrt{3}}(\eta^* |0\rangle+ |1\rangle+ |2\rangle),\nonumber \\ 
    |Z \rangle_1 &=\frac{1}{\sqrt{3}}(|0\rangle+\eta^* |1\rangle+|2\rangle),\\
    |Z \rangle_2 &=\frac{1}{\sqrt{3}}(|0\rangle+|1\rangle+\eta^* |2\rangle).\nonumber
\end{align}

Following the same procedure as in the previous subsection, we obtain the following expression for $t$ and the other probabilities
\begin{align}
   t &=1-\frac{1}{6}(P_{X_0X_1}+P_{X_0X_2}+P_{X_1X_0}+P_{X_1X_2} \nonumber \\
& \quad +P_{X_2X_1}+P_{X_2X_0}+P_{Y_0Y_1}+P_{Y_0Y_2}+P_{Y_1Y_0} \\
& \quad +P_{Y_1Y_2}+P_{Y_2Y_1}+P_{Y_2Y_0}+P_{Z_0Z_1}+P_{Z_0Z_2}\nonumber \\
& \quad +P_{Z_1Z_0}+P_{Z_1Z_2}+P_{Z_2Z_1}+P_{Z_2Z_0}),\nonumber
\end{align}
where, just like before, the probabilities $P_{ij}$ for $\left\{i,j \right\}= \left\{X_0, X_1, X_2\right\}$, $\left\{i,j \right\}= \left\{Y_0, Y_1, Y_2\right\}$ or $\left\{i,j \right\}= \left\{Z_0, Z_1, Z_2\right\}$, stand for the cases where Alice measures the returned qutrit in the state $|j\rangle$ when she originally prepared it in the state $|i\rangle$. We can also notice that the remaining other scalar products are zero, \textit{i.e.} $a=b=c=z=m=0$. In this case, we obtain the following bound

\begin{equation}
\label{X4MUBS}
\begin{split}
    X_{4MUBs}  \geq & 3-\frac{1}{2}(P_{X_0X_1}+P_{X_0X_2}+P_{X_1X_0}+P_{X_1X_2}\\
    & \quad +P_{X_2X_1}+P_{X_2X_0}+P_{Y_0Y_1}+P_{Y_0Y_2}+P_{Y_1Y_0} \\
    & \quad +P_{Y_1Y_2}+P_{Y_2Y_1}+P_{Y_2Y_0}+P_{Z_0Z_1}+P_{Z_0Z_2}\\
    &\quad  +P_{Z_1Z_0}+P_{Z_1Z_2}+P_{Z_2Z_1}+P_{Z_2Z_0}) \\
&-3(\sqrt {p_{000} p_{101}}+\sqrt {p_{010} p_{101}}+\sqrt {p_{020} p_{101}}\\
&\quad +\sqrt {p_{010} p_{111}}+\sqrt {p_{020} p_{111}} +\sqrt {p_{000} p_{121}}\\
&+\sqrt {p_{010} p_{121}}+\sqrt {p_{020} p_{121}}).
\end{split}
\end{equation}

We introduce the following notation
\begin{equation}
    \mathcal{S} = \begin{cases} \phantom{-} X_{i}^2& \text{if } X_{i} 
\geq 0 \\ \phantom{-} 0 & otherwise \end{cases},
\end{equation}
where $i=\left\{3MUBs,4MUBs\right\}$, thus combining the two cases studied here
\begin{eqnarray}
    X_{i}=Re(\langle e^0_{0,00}|e^1_{1,11}\rangle)+Re(\langle e^0_{0,00}|e^2_{2,22}\rangle)+Re(\langle e^1_{1,11}|e^2_{2,22}\rangle).
\end{eqnarray}

Using the lower bound, on this last quantity, derived in (\ref{X3MUBS}, \ref{X4MUBS}) for three and four MUBs respectively and the fact that $p=|\langle e^0_{0,0}|e^1_{1,4}\rangle+\langle e^0_{0,0}|e^2_{2,8}\rangle+\langle e^1_{1,4}|e^2_{2,8}\rangle|^2 \geq \mathcal{S}$, allows to lower bound $p$ thus upper bound the von Neumann entropy $S(EC)$ and ultimately allow to find a lower bound on the conditional entropy $S(B|EC)$ in (\ref{strongsubadditivity}).

\begin{figure}[htbp]
    \hspace{-4mm}
    \begin{minipage}{0.5\linewidth}
        \centering
        \includegraphics[width=\textwidth]{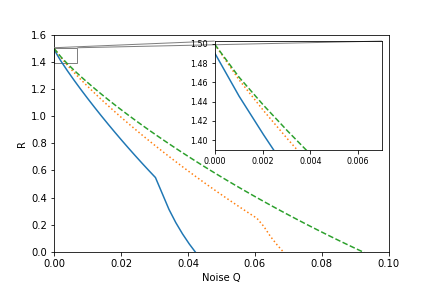}\\
        (a)
    \end{minipage}
    \begin{minipage}{0.5\linewidth}
        \centering
        \includegraphics[width=\textwidth]{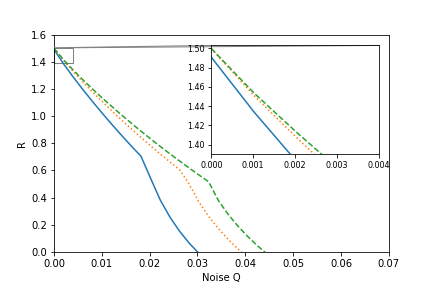}\\
         (b)
    \end{minipage}
    \caption{The key rate as a function of the noise $Q$ for three-dimensional quantum states with two(blue line), three(orange line), and four(green line) mutually unbiased bases. (a)  The dependent channel case, (b) The independent channel case }
     \label{fig:3D}
\end{figure}

Figure \ref{fig:3D} shows a plot of the numerical solutions for the lower bound of the key rate $r$ as a function of the noise $Q$ in the two scenarios, \textit{dependent channel} (Fig.\ref{fig:3D}a) and \textit{independent channel} (Fig.\ref{fig:3D}b). Our results clearly show that the threshold value of tolerated noise increases as the number of the mutually unbiased bases are increased, but the key rate itself increases weakly for small values of the noise. 
These results are summarised in Table \ref{table3D}.

\begin{table}[!t]
\renewcommand{\arraystretch}{1.3}
\centering % used for centering table
\caption{Maximal noise tolerance for a three-dimensional SQKD protocol, in a variety of scenarios.} 
\begin{tabular}{c c c } % centered columns (4 columns)
\hline\hline %inserts double horizontal lines
Number of MUBs used & Dependent & Independent  \\ [0.5ex] % inserts table
%heading
\hline % inserts single horizontal line
 2 MUBs\cite{hajji2021qutrit} & $4.247\%$ & $3.05 \%$ \\ % inserting body of the table
 3 MUBs & $6.89\%$ & $3.95\%$  \\
 4 MUBs & $9.32\%$ & $4.43\%$  \\ [1ex] % [1ex] adds vertical space
\hline %inserts single line
\end{tabular}
\label{table3D} 
\end{table}

\section{COLLECTIVE EAVESDROPPING WITH \\
FOUR-DIMENSIONAL QUANTUM STATES}

We will now consider the case of SQKD protocol based on ququart instead of qutrits \textit{i.e.} four dimensional systems $d=4$. We derive the security against collective eavesdropping strategy when the protocol uses  two, three, four or five mutually unbiased bases. 

The computational basis being $\{ |0\rangle,|1\rangle,|2\rangle,|3\rangle\}$, one can write the most general unitary eavesdropping strategy for a set of four-dimensional quantum states, in the forward channel as

\setlength{\arraycolsep}{0.0em}
\begin{eqnarray}
|0\rangle\otimes |E\rangle &\xrightarrow{U_F}&|0\rangle|e_{00}\rangle +|1\rangle|e_{01}\rangle+|2\rangle|e_{02}\rangle+|3\rangle|e_{03}\rangle,  \nonumber \\
|1\rangle\otimes |E\rangle &\xrightarrow{U_F}&|0\rangle|e_{10}\rangle  +|1\rangle|e_{11}\rangle+|2\rangle|e_{12}\rangle+|3\rangle|e_{13}\rangle,\nonumber \\
|2\rangle\otimes |E\rangle &\xrightarrow{U_F}&|0\rangle|e_{20}\rangle  +|1\rangle|e_{21}\rangle+|2\rangle|e_{22}\rangle+|3\rangle|e_{23}\rangle,  \\
|3\rangle\otimes |E\rangle &\xrightarrow{U_F}&|0\rangle|e_{30}\rangle  +|1\rangle|e_{31}\rangle+|2\rangle|e_{32}\rangle+|3\rangle|e_{33}\rangle. \nonumber
\end{eqnarray}
\setlength{\arraycolsep}{5pt}

The attack on the backward channel will yield the final global state as follows
\begin{equation}
\label{ur}
    |i, e_j\rangle \xrightarrow{U_R} |0, e^0_{i,j} \rangle + |1, e^1_{i,j}\rangle+ |2, e^2_{i,j}\rangle+ |3, e^3_{i,j}\rangle,
\end{equation}

We follow the same procedure as in the previous section and start by consider only the process of the protocol in which an iteration is used to contribute towards the raw key, \textit{i.e.} instances where Alice sends a computational basis state (either $|0\rangle$, $|1\rangle$ , $|2\rangle$ or $|3\rangle$ each chosen with probability 1/4), Bob measures it, then re-sends it.

The strong sub-additivity of the Von Neumann entropy, allows to derive the lower bound on the key rate:
\setlength{\arraycolsep}{0.0em}
\begin{eqnarray}
    r &\geq& H(\frac{1}{4} p_{0,0,0},\frac{1}{4} p_{0,0,1},..., \frac{1}{4} p_{3,3,3})- H\left(\frac{1}{4}t_1,\frac{1}{4}t_2,\frac{1}{4}t_3, \frac{1}{4}t_4\right) \nonumber \\ &-&\frac{1}{4}(t_2+t_3+t_4)-\frac{1}{4}t_1H(\Tilde{\lambda_1},\Tilde{\lambda_2} )\\
    &+&H(p_A(0),p_A(1),p_A(2),p_A(3))- H(\left\{ p(i,j)\right\}_{i,j=0,1,2,3}  ).\nonumber
\end{eqnarray}
where we have used the same notations as in (\ref{keyratebound3}) and the eigenvalues $\Tilde{\lambda_i}$ are given by

\begin{align}
\label{lambdatilde4}
\Tilde{\lambda_1}&=\frac{1}{2}+\frac{\sqrt{4p-4p_{1,1,1}p_{2,2,2}-4p_{0,0,0}p_{3,3,3}+(p_{0,0,0}-p_{1,1,1}-p_{2,2,2}+p_{3,3,3})^2}}{2(p_{0,0,0}+p_{1,1,1}+p_{2,2,2}+p_{3,3,3})}, \\  %\label{lambda1tilde4} \\
\Tilde{\lambda_2}&=\frac{1}{2}-\frac{\sqrt{4p-4p_{1,1,1}p_{2,2,2}-4p_{0,0,0}p_{3,3,3}+(p_{0,0,0}-p_{1,1,1}-p_{2,2,2}+p_{3,3,3})^2}}{2(p_{0,0,0}+p_{1,1,1}+p_{2,2,2}+p_{3,3,3})},\nonumber  %\label{lambda2tilde4}\\
\end{align}
where $p=\left(|\langle e^0_{0,00}|e^1_{1,11}\rangle|^2+|\langle e^0_{0,00}|e^2_{2,22}\rangle|^2+|\langle e^0_{0,00}|e^3_{3,33}\rangle|^2
+|\langle e^1_{1,11}|e^2_{2,22}\rangle|^2+|\langle e^1_{1,11}|e^3_{3,33}\rangle|^2+|\langle e^2_{2,22}|e^3_{3,33}\rangle|^2\right)$.\\

A lower bound of the quantity $p$ by using the error rate in the different numbers of mutually unbiased bases is needed in order to derive an expression of a lower bound of $r$. We will carry this next, depending on the number of bases used in the protocol. In fact for a four dimensional Hilbert space one can define up to five mutually unbiased bases \cite{karol}. So in addition to the computational basis $\left\{ |i\rangle \right\}_{i=0,1,2,3} $ we can add four other bases. Conventionally, one chooses the first basis to be the discrete Fourier transform to the computational basis 
\begin{align}
\label{Abasis}
    |A\rangle_0 &=\frac{1}{2}(|0\rangle+ |1\rangle+ |2\rangle+|3\rangle), \nonumber \\
    |A\rangle_1 &=\frac{1}{2}(|0\rangle+ |1\rangle- |2\rangle-|3\rangle),\nonumber \\
    |A\rangle_2 &=\frac{1}{2}(|0\rangle- |1\rangle- |2\rangle+|3\rangle),\\
    |A\rangle_3 &=\frac{1}{2}(|0\rangle- |1\rangle+ |2\rangle-|3\rangle).\nonumber
\end{align}

The second basis is defined as 
\begin{align}
\label{Bbasis}
   |B\rangle_0 &=\frac{1}{2}(|0\rangle- |1\rangle-i |2\rangle-i |3\rangle),\nonumber\\
   |B\rangle_1 &=\frac{1}{2}(|0\rangle- |1\rangle+i |2\rangle+i |3\rangle),\\
   |B\rangle_2 &=\frac{1}{2}(|0\rangle+ |1\rangle+i |2\rangle-i |3\rangle),\nonumber \\
   |B\rangle_3 &=\frac{1}{2}(|0\rangle+ |1\rangle-i |2\rangle+i |3\rangle).\nonumber
\end{align}
Similarly, the third basis is defined as
\begin{align}
\label{Cbasis}
   |C\rangle_0 &=\frac{1}{2}(|0\rangle-i |1\rangle-i  |2\rangle- |3\rangle),\nonumber\\
   |C\rangle_1 &=\frac{1}{2}(|0\rangle-i |1\rangle+i  |2\rangle+|3\rangle),\nonumber\\
   |C\rangle_2 &=\frac{1}{2}(|0\rangle+i |1\rangle+i  |2\rangle-|3\rangle),\\
   |C\rangle_3 &=\frac{1}{2}(|0\rangle+i  |1\rangle-i  |2\rangle+|3\rangle),\nonumber
\end{align}
while the fourth basis is defined as
\begin{align}
\label{Dbasis}
  |D\rangle_0 &=\frac{1}{2}(|0\rangle-i |1\rangle- |2\rangle-i |3\rangle), \nonumber\\
  |D\rangle_1 &=\frac{1}{2}(|0\rangle-i |1\rangle+ |2\rangle+i |3\rangle),\nonumber\\
  |D\rangle_2 &=\frac{1}{2}(|0\rangle+i |1\rangle- |2\rangle+i |3\rangle),\\
  |D\rangle_3 &=\frac{1}{2}(|0\rangle+i|1\rangle+ |2\rangle-i |3\rangle).\nonumber
\end{align}
The above states represent the maximum number of mutually unbiased bases for ququarts.

When considering those iterations where Alice initially sends, then measures in these bases depending on the numbers of MUBs, while Bob chooses to reflect the ququart, Bob's operation is essentially the identity operator while Eve's action is again the unitary operation $V = U_RU_F$:
\setlength{\arraycolsep}{0.0em}
\begin{equation}
   |i\rangle\otimes |0\rangle_{E} \xrightarrow{V} |i\rangle | f_{ii} \rangle + |i+1\rangle | f_{ii+1}\rangle+ |i+2\rangle | f_{ii+2}\rangle+|i+3\rangle | f_{ii+3}\rangle,
\end{equation}
where $i=0,1,2,3$ and the subscript addition is taken modulo 4.

The scalar products between Eve's output states have to obey similar constraints to those for the  three-dimensional case (\ref{UniV3}) leading again to classification of Eve's output states into six sets of scalar products, each defining a free parameter:
\begin{align}
a=&\langle f_{ii}|f_{ij}\rangle , &&\text{for $i\neq j$} , \nonumber \\
b=&\langle f_{ii}|f_{jk}\rangle , &&\text{where}\;  i,j, \text{and}\; k \; \text{are all different,}\nonumber\\
c=&\langle f_{ij}|f_{ik}\rangle ,  &&\text{where}\;  i,j, \text{and}\; k \; \text{are all different,}\nonumber\\
z=&\langle f_{ij}|f_{jh}\rangle ,  &&\text{where}\;  i,j, \text{and}\; h \; \text{are all different,}\\
m=&\langle f_{ij}|f_{hk}\rangle ,  &&\text{where} \; j\neq i, (h=j \; \text{and}\; k=i) \; \nonumber \\ 
&&& \text{or}\; (h, k, i \; \text{and}\; j \; \text{are all different}); \; \text{m is real,} \nonumber \\
t=&\langle f_{ii}|f_{jj}\rangle ,  &&\text{for $i\neq j$; t is also real.}\nonumber
\end{align}

Combining the symmetry condition, unitary condition and the errors Eve's attack induces when using different numbers of MUBs, we can derive the expression summarized in Table \ref{table:nonlin}.
\begin{table}[!t]
\caption{Results of estimating the noise in ququart based SQKD protocols using different numbers of MUBs. We use the notations $\displaystyle A= \sum_{i,j; i\neq j}{p_{ij}}$ for $\left\{i,j \right\}= \left\{A_0, A_1, A_2, A_3 \right\}$, $\displaystyle B= \sum_{i,j; i\neq j}{p_{ij}}$ for $\left\{i,j \right\}= \left\{B_0, B_1, B_2, B_3 \right\}$, $ \displaystyle C= \sum_{i,j; i\neq j}{p_{ij}}$ for $\left\{i,j \right\}= \left\{C_0, C_1, C_2, C_3 \right\}$, $\displaystyle D= \sum_{i,j; i\neq j}{p_{ij}}$ for $\left\{i,j \right\}= \left\{D_0, D_1, D_2, D_3 \right\}$, to denote the probability of the event that Alice measures the returned ququart in the state $|j\rangle$ when she originally prepared it in the state $|i\rangle$ in the corresponding basis.}
\centering % used for centering table
\begin{tabular}{c c c } % centered columns (4 columns)
\hline\hline %inserts double horizontal lines
Bases & Vectors & Expression  \\ [0.5ex] % inserts table
%heading
\hline % inserts single horizontal line
2 & 4 & $t=1-\frac{1}{3}A-3Re(m)$ \\ % inserting body of the table
3 & 8 & $t=1-\frac{1}{6}(A+B)-Re(m)$  \\
4 & 12 & $t=1-\frac{1}{9}(A+B+C)-\frac{1}{3}Re(m)$  \\
5 & 16 & $t=1-\frac{1}{12}(A+B+C+D)$  \\ [1ex] % [1ex] adds vertical space
\hline %inserts single line
\end{tabular}
\label{table:nonlin} % is used to refer this table in the text
\end{table} 

From Cauchy-Schwarz' inequality one can bound the quantity $ W_{i-MUBs}=Re(p)$ as follows
\begin{itemize}
   \item[$\square$] \textit{Two MUBs}:
        \begin{equation}
           \begin{split}
    W_{2-MUBs}  \geq & 6-2(P_{A_0A_1}+P_{A_0A_2}+P_{A_0A_3}+P_{A_1A_0}+P_{A_1A_2}+P_{A_1A_3}\\
      & \quad +P_{A_2A_1}+P_{A_2A_0}+P_{A_2A_3}+P_{A_3A_1}+P_{A_3A_0}+P_{A_3A_2})\\
& \quad  -18(\sqrt{P_{100}P_{001}} + \sqrt{P_{110}P_{001}} + \sqrt{P_{120}P_{001}}+\sqrt{P_{130}P_{001}} \\
& \quad  + \sqrt{P_{100}P_{011}} + \sqrt{P_{110}P_{011}} + \sqrt{P_{120}P_{011}} + \sqrt{P_{130}P_{011}} + \sqrt{P_{100}P_{021}} \\
& \quad + \sqrt{P_{110}P_{021}} + \sqrt{P_{120}P_{021}} + \sqrt{P_{130}P_{021}} + \sqrt{P_{100}P_{031}} + \sqrt{P_{110}P_{031}} \\
 & \quad + \sqrt{P_{120}P_{031}} + \sqrt{P_{130}P_{031}})-6(\sqrt{P_{101}P_{000}} +  \sqrt{P_{121}P_{000}} + \sqrt{P_{131}P_{000}} \\
 & \quad + \sqrt{P_{101}P_{010}} + \sqrt{P_{111}P_{010}} +\sqrt{P_{121}P_{010}} + \sqrt{P_{131}P_{010}} + \sqrt{P_{101}P_{020}} \\
 & \quad + \sqrt{P_{111}P_{020}} + \sqrt{P_{121}P_{020}} + \sqrt{P_{131}P_{020}} + \sqrt{P_{101}P_{030}} + \sqrt{P_{111}P_{030}} \\
 & \quad + \sqrt{P_{121}P_{030}} + \sqrt{P_{131}P_{030}}).
\end{split}
\end{equation}
    \item [$\square$] \textit{Three MUBs}:
\begin{equation}
\begin{split}
    W_{3-MUBs}  \geq & 6-(P_{A_0A_1}+P_{A_0A_2}+P_{A_0A_3}+P_{A_1A_0}+P_{A_1A_2}+P_{A_1A_3}\\
      & \quad +P_{A_2A_1}+P_{A_2A_0}+P_{A_2A_3}+P_{A_3A_1}+P_{A_3A_0}+P_{A_3A_2}\\
       & \quad +P_{B_0B_1}+P_{B_0B_2}+P_{B_0B_3}+P_{B_1B_0}+P_{B_1B_2}+P_{B_1B_3}\\
      & \quad +P_{B_2B_1}+P_{B_2B_0}+P_{B_2B_3}+P_{B_3B_1}+P_{B_3B_0}+P_{B_3B_2})\\
& \quad  -6(\sqrt{P_{100}P_{001}} + \sqrt{P_{110}P_{001}} + \sqrt{P_{120}P_{001}}+\sqrt{P_{130}P_{001}} \\
& \quad  + \sqrt{P_{100}P_{011}} + \sqrt{P_{110}P_{011}} + \sqrt{P_{120}P_{011}} + \sqrt{P_{130}P_{011}} + \sqrt{P_{100}P_{021}} \\
& \quad + \sqrt{P_{110}P_{021}} + \sqrt{P_{120}P_{021}} + \sqrt{P_{130}P_{021}} + \sqrt{P_{100}P_{031}} + \sqrt{P_{110}P_{031}} \\
 & \quad + \sqrt{P_{120}P_{031}} + \sqrt{P_{130}P_{031}})-6(\sqrt{P_{101}P_{000}} +  \sqrt{P_{121}P_{000}} + \sqrt{P_{131}P_{000}} \\
 & \quad + \sqrt{P_{101}P_{010}} + \sqrt{P_{111}P_{010}} +\sqrt{P_{121}P_{010}} + \sqrt{P_{131}P_{010}} + \sqrt{P_{101}P_{020}} \\
 & \quad + \sqrt{P_{111}P_{020}} + \sqrt{P_{121}P_{020}} + \sqrt{P_{131}P_{020}} + \sqrt{P_{101}P_{030}} + \sqrt{P_{111}P_{030}} \\
 & \quad + \sqrt{P_{121}P_{030}} + \sqrt{P_{131}P_{030}}).
\end{split}
\end{equation}
     \item [$\square$] \textit{Four MUBs}:
\begin{equation}
\begin{split}
    W_{4-MUBs}  \geq & 6-\frac{2}{3}(P_{A_0A_1}+P_{A_0A_2}+P_{A_0A_3}+P_{A_1A_0}+P_{A_1A_2}+P_{A_1A_3}\\
      & \quad +P_{A_2A_1}+P_{A_2A_0}+P_{A_2A_3}+P_{A_3A_1}+P_{A_3A_0}+P_{A_3A_2}\\
       & \quad +P_{B_0B_1}+P_{B_0B_2}+P_{B_0B_3}+P_{B_1B_0}+P_{B_1B_2}+P_{B_1B_3}\\
      & \quad +P_{B_2B_1}+P_{B_2B_0}+P_{B_2B_3}+P_{B_3B_1}+P_{B_3B_0}+P_{B_3B_2}\\
      & \quad +P_{C_0C_1}+P_{C_0C_2}+P_{C_0C_3}+P_{C_1C_0}+P_{C_1C_2}+P_{C_1C_3}\\
      & \quad +P_{C_2C_1}+P_{C_2C_0}+P_{C_2C_3}+P_{C_3C_1}+P_{C_3C_0}+P_{C_3C_2})\\
& \quad  -2(\sqrt{P_{100}P_{001}} + \sqrt{P_{110}P_{001}} + \sqrt{P_{120}P_{001}}+\sqrt{P_{130}P_{001}} \\
& \quad  + \sqrt{P_{100}P_{011}} + \sqrt{P_{110}P_{011}} + \sqrt{P_{120}P_{011}} + \sqrt{P_{130}P_{011}} + \sqrt{P_{100}P_{021}} \\
& \quad + \sqrt{P_{110}P_{021}} + \sqrt{P_{120}P_{021}} + \sqrt{P_{130}P_{021}} + \sqrt{P_{100}P_{031}} + \sqrt{P_{110}P_{031}} \\
 & \quad + \sqrt{P_{120}P_{031}} + \sqrt{P_{130}P_{031}})-6(\sqrt{P_{101}P_{000}} +  \sqrt{P_{121}P_{000}} + \sqrt{P_{131}P_{000}} \\
 & \quad + \sqrt{P_{101}P_{010}} + \sqrt{P_{111}P_{010}} +\sqrt{P_{121}P_{010}} + \sqrt{P_{131}P_{010}} + \sqrt{P_{101}P_{020}} \\
 & \quad + \sqrt{P_{111}P_{020}} + \sqrt{P_{121}P_{020}} + \sqrt{P_{131}P_{020}} + \sqrt{P_{101}P_{030}} + \sqrt{P_{111}P_{030}} \\
 & \quad + \sqrt{P_{121}P_{030}} + \sqrt{P_{131}P_{030}}).
\end{split}
\end{equation}
      \item [$\square$] \textit{Five MUBs}:
\begin{equation}
\begin{split}
    W_{5-MUBs}  \geq & 6-\frac{1}{2}(P_{A_0A_1}+P_{A_0A_2}+P_{A_0A_3}+P_{A_1A_0}+P_{A_1A_2}+P_{A_1A_3}\\
      & \quad +P_{A_2A_1}+P_{A_2A_0}+P_{A_2A_3}+P_{A_3A_1}+P_{A_3A_0}+P_{A_3A_2}\\
       & \quad +P_{B_0B_1}+P_{B_0B_2}+P_{B_0B_3}+P_{B_1B_0}+P_{B_1B_2}+P_{B_1B_3}\\
      & \quad +P_{B_2B_1}+P_{B_2B_0}+P_{B_2B_3}+P_{B_3B_1}+P_{B_3B_0}+P_{B_3B_2}\\
      & \quad +P_{C_0C_1}+P_{C_0C_2}+P_{C_0C_3}+P_{C_1C_0}+P_{C_1C_2}+P_{C_1C_3}\\
      & \quad +P_{C_2C_1}+P_{C_2C_0}+P_{C_2C_3}+P_{C_3C_1}+P_{C_3C_0}+P_{C_3C_2}\\
      & \quad +P_{D_0D_1}+P_{D_0D_2}+P_{D_0D_3}+P_{D_1D_0}+P_{D_1D_2}+P_{D_1D_3}\\
      & \quad +P_{D_2D_1}+P_{D_2D_0}+P_{D_2D_3}+P_{D_3D_1}+P_{D_3D_0}+P_{D_3D_2})\\
& \quad  -6(\sqrt{P_{101}P_{000}} +  \sqrt{P_{121}P_{000}} + \sqrt{P_{131}P_{000}} \\
 & \quad + \sqrt{P_{101}P_{010}} + \sqrt{P_{111}P_{010}} +\sqrt{P_{121}P_{010}} + \sqrt{P_{131}P_{010}} + \sqrt{P_{101}P_{020}} \\
 & \quad + \sqrt{P_{111}P_{020}} + \sqrt{P_{121}P_{020}} + \sqrt{P_{131}P_{020}} + \sqrt{P_{101}P_{030}} + \sqrt{P_{111}P_{030}} \\
 & \quad + \sqrt{P_{121}P_{030}} + \sqrt{P_{131}P_{030}}).
\end{split}
\end{equation}
\end{itemize}

Again, making use of the  following notation
\begin{equation}
    \mathcal{S} = \begin{cases} \phantom{-} W_{i}^2& \text{if } W_{i} 
\geq 0 \\ \phantom{-} 0 & otherwise \end{cases},
\end{equation}
where $i=\left\{2MUBs,3MUBs,4MUBs,5MUBs\right\}$ and 
\begin{eqnarray}
    W_{i}=&& Re(\langle e^0_{0,00}|e^1_{1,11}\rangle)+Re(\langle e^0_{0,00}|e^2_{2,22}\rangle)+Re(\langle e^0_{0,00}|e^3_{3,33}\rangle)\nonumber \\ && +Re(\langle e^1_{1,11}|e^2_{2,22}\rangle)+Re(\langle e^1_{1,11}|e^3_{3,33}\rangle)+Re(\langle e^2_{2,22}|e^3_{3,33}\rangle),
\end{eqnarray}

The bound on this quantity allows to easily find a lower bound on the key rate which is expressed as a function of parameters determined by the quantum channel. Thus we can compute the intuitive result that the secret key rate is positive if Eve has less information than Bob.

\begin{figure}[htbp]
    \hspace{-4mm}
    \begin{minipage}{0.5\linewidth}
        \centering
        \includegraphics[width=\textwidth]{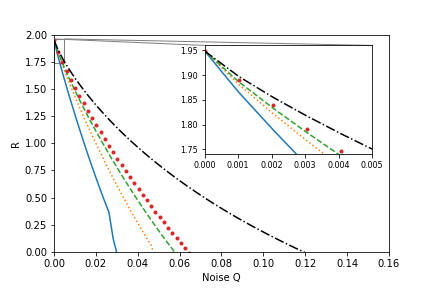}\\
        (a)
    \end{minipage}
    \begin{minipage}{0.5\linewidth}
        \centering
        \includegraphics[width=\textwidth]{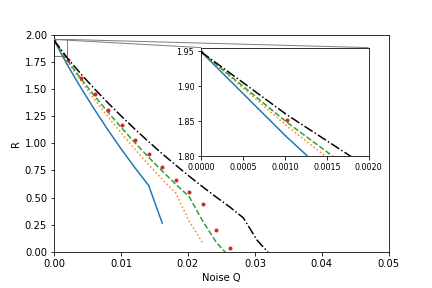}\\
       (b)
    \end{minipage}
        \caption{The key rate as a function of the noise Q for four-dimensional quantum states with two (blue line), three (orange line), four (green line), and five (red line) mutually unbiased bases. (a) shows the dependent channel ( when the $\left\{ |k\rangle \right\}_{i}$ for ${k=A,B,C,D},{i=0,1,2,3} $ basis noise is $Q$) whereas (b) represents the independent channel (in this case the noise is $Q_{k}$=2Q(3-6Q) for ${k=A,B,C,D}$).}
    \label{fig4D}
\end{figure}

The lower bound of the key rate as a function of the noise parameter $Q$, for various numbers of mutually unbiased bases, are plotted in Figure \ref{fig4D}. We consider two forms in particular: dependent (Figure \ref{fig4D}a) and independent channels (Figure \ref{fig4D}b). We observe numerically that, as the number of MUBs increases, the maximal noise tolerance goes up to $6,48\%$ in the dependent case and to $2,65\%$ in the independent case both obtained when using the maximum number of MUBs \textit{i.e.} four MUBs. One can also observe by comparing these results to the results in the previous section, when using qutrits, that the noise tolerance for four the dimensional scheme is lower compared to the three dimensional scheme but the reward being in increase in the key generation rate, which goes from $1.5$ to $2$.

To show the importance of the choice of the bases to work with, once the choice of the number is made, we show in the same figures (Fig. \ref{fig4D}) the case when the two mutually unbiased bases, namely the computational basis and the second base (\ref{Bbasis}) is being used. This is shown as the black line in these figures. It turns out that the maximal noise tolerance in the dependent channel is about $12.05\%$, and it drops to $3.22\%$ in the independent channel. This shows that in some sens that  qualitative choice is also important not only a quantitative one \textit{i.e.} not only the number of MUBs used is important but also which bases are chosen.

The different results, about the maximum noise tolerated, are summarized in Table \ref{table4D}.

\begin{table}%[!t]
\renewcommand{\arraystretch}{1.3}
\centering % used for centering table
\caption{Maximal noise tolerance for a four-dimensional SQKD protocol under both dependent and independent channel scenarios.}
\begin{tabular}{c c c } % centered columns (4 columns)
\hline\hline %inserts double horizontal lines
Number of MUBs used & Dependent & Independent  \\ [0.5ex] % inserts table
%heading
\hline % inserts single horizontal line
 2MUBs & $3\%$ & $1.62\%$ \\ % inserting body of the table
 3MUBs & $4.77\%$ & $2.24\%$  \\
 4MUBs & $5.79\%$ & $2.58\%$  \\ 
 5MUBs & $6.48\%$ & $2.65\%$  \\ [1ex] % [1ex] adds vertical space
\hline %inserts single line
\end{tabular}
\label{table4D} % is used to refer this table in the text
\end{table}

\section {CONCLUSIONS}
Semi-quantum key distribution protocols can be formulated and implemented using systems with different dimensions (qubits, qutrits...). Once the dimension is increased, the number of mutually unbiased bases upon which to base the protocol opens leading to numerous versions and variants of the same protocol. Here we have discussed the robustness of qutrit and ququart based semi-quantum key distribution protocols with different mutually unbiased bases, under the assumption of collective eavesdropping attacks. We have derived a lower bound on the key rate, as a function only of the quantum channel’s noise (a parameter that may be estimated by the legitimate parties), and evaluated its lower bound by considering the most common forms of channels, namely the independent channel and dependent channel.

The main findings are:
\begin{enumerate}
    \item The increase in the number of mutually unbiased bases, leads to an increase in the quantum channel’s noise tolerated.
    \item Encoding in higher-dimensional MUBs, improve the robustness against eavesdropping and allows for a higher generation rates of secure key while the price is a lower error rate.
\end{enumerate}

Meanwhile, a qualitative analysis suggests that in some cases choosing to work with some two bases can be preferred to working with $d+1$ bases since as the  maximum error rate is much larger.

\section*{Acknowledgements}

This work has been supported by the National Center for Scientific and Technical Research (CNRST). 

%\section*{References}
\bibliographystyle{abbrv}
\bibliography{MUBSSQKD}

\end{document}